\documentclass[aps,showpacs,nofootinbib]{revtex4}
\usepackage{amsmath} 
\usepackage{amssymb}
\usepackage{graphicx} 
\usepackage{color} 
\usepackage{epstopdf}

\begin{document}

\newcommand{\blue}[1]{\textcolor{blue}{#1}}
\newcommand{\new}{\blue}
\newcommand{\green}[1]{\textcolor{green}{#1}}
\newcommand{\modif}{\green}
\newcommand{\red}[1]{\textcolor{red}{#1}}
\newcommand{\attention}{\red}


\title{Magnetic field in nuclear collisions at ultra high energies}

\author{V. A. Okorokov} \email{VAOkorokov@mephi.ru; Okorokov@bnl.gov}
\affiliation{National Research Nuclear University MEPhI (Moscow
Engineering Physics Institute), Kashirskoe highway 31, 115409
Moscow, Russia}

\date{\today}

\begin{abstract}
The magnetic field created in proton-proton and nucleus--nucleus
collisions at ultra high energies are studied with models of
point-like charges and hard sphere for distribution of the
constituents for vacuum conditions. The various
beam ions are considered from light to heavy nuclei at energies
corresponded to the nominal energies of proton beam within the
projects of further accelerator facilities high--energy Large Hadron Collider (HE--LHC) and
Future Circular Collider (FCC). The
magnetic field strength immediately after collisions reaches the
value tens of GeV$^{2}$ while the approach with point-like charges
some overestimate the amplitude of the field in comparison with
more realistic hard sphere model. The absolute value of magnetic
field rapidly decrease with time and increases with growth of
atomic number. The amplitude for $eB$ is estimated at level $100$
GeV$^{2}$ in order to magnitude for quark-quark collisions at
energies corresponded to the nominal energies of proton beams.
These estimations are close to the range for onset of $W$ boson
condensation.
\end{abstract}


\maketitle

\section{Introduction}\label{sec:1}

According to the Biot--Savart law the current, i.e. moving charge,
create magnetic field ($\vec{B}$). Therefore the collisions of
charged particles in human made accelerator facilities or in
cosmic rays generate the magnetic field which strength can
achieves very large value. This field appears just after collision
moment and, consequently, can influence on all stages of
space-time evolution of final-state system. The influence of
$\vec{B}$ and corresponding electric filed ($\vec{E}$) can be
essential for phase diagram of the matter created in final state
and for transition processes at sufficiently large strength of
this external Abelian (electro)magnetic field. Also it can lead to
some new features for dynamics of multiparticle production. In
general the maximum of the absolute value $B \equiv |\vec{B}|$
will increase with growth of the energy of incoming particles and,
consequently, it can be expected the amplification of the
influence of (electro)magnetic field on the various properties of
the final state in the domain of very high energies. As
consequence, the study of possible influence of external Abelian
(electro)magnetic field created in the collisions of relativistic
particles on the interaction process is important for both the
strong interaction and the electroweak sector. Collisions of
particles with ultra relativistic energies provide a unique
possibility to study the wide set of physical
effects related to the very strong electromagnetic fields at controlled conditions. On the other
hand, such investigations can shed new light on the physical
mechanisms that may have produced magnetic fields in the early
Universe. Therefore, studies of the extremely strong
electromagnetic field in particle collisions with ultra high
energies can be important for physics of fundamental interactions,
cosmology, and relativistic astrophysics, i.e., they have an
interdisciplinary value.

\section{\label{sec:2}Definitions and notations}

In this section some essential kinematic parameters for collider beams and approach for magnetic field study are considered.

\subsection{\label{subsec:2.1}Models for magnetic field}
In the simplest approach one can assume like in
\cite{PLB-377-135-1996} that the colliding objects are point-like
particles with charges $\mbox{Z}e$, where $e \equiv |e|$ is the
magnitude of electron charge. Lets the objects move along the $Z$
axis at impact parameter $\vec{b}$. Then time evolution of $B$ at
the center of the collision can be described by the following
equation \cite{PRD-80-034028-2009}
\begin{equation}
B(t)=B_{0}\bigl[1+(t/t_{0})^{3/2}\bigr]^{-1},~~~ eB_{0}=8 \mbox{Z}\alpha_{\mbox{\scriptsize{EM}}}\sinh(y_{0})/b^{2},~~~ t_{0}=b/[2\sinh(y_{0})].\label{eq:2.0}
\end{equation}
Here $\alpha_{\mbox{\scriptsize{EM}}}$ is the electromagnetic
constant and $y_{0}$ is the rapidity of incoming particles in the
center-of-mass system. In relativistic energy domain ($p_{\,0} \gg
m$) the relations $\sinh(y_{0})=\gamma_{0}\beta_{0,z} \approx
\sqrt{s}/2m$ result in
$$
eB_{0} \propto \sqrt{s},
$$
where $p_{\,0}$, $\gamma_{0}$, $\beta_{0,z}$ and $m$ is the
momentum, Lorentz factor, velocity along the $Z$ axis, and mass of
incoming particle in the frame considered; $s$ is standard
Mandelstam invariant variable.

Magnetic field at a point
$\vec{x}=(\vec{x}_{\perp}, z\vec{e}_{\scriptsize{Z}})$ created by an object (proton / nucleus) with finite size and a charge $\mbox{Z}e$,
moving in the positive ($z>0$) direction of the $Z$ axis from
the starting point of the transverse plane
$\vec{x}^{\,'}_{\perp}=\vec{x}_{\perp}\left|_{t = 0}\right.$, can
be obtained either with help of appropriate conversation of electric
field of a given charged object, or on the basis of the Li$\acute{\mbox{e}}$nard--Wihert potentials \cite{NPA-803-227-2008}.
In this work the magnetic field created immediately after the collision is studied, i.e. $\vec{B}$ at times $t>0$, where $t=0$ is the collision moment. Therefore $\vec{B}$ can be written as follows \cite{NPA-803-227-2008}:
\begin {equation}
\vec{B} = \sum \limits_{i = +, -}\sum \limits_{j = \mbox{\scriptsize{s, p}}} \vec{B}^{i}_{j},
\label{eq:2.1}
\end{equation}
where $\vec{B}^{\pm}_{\mbox{\scriptsize{s}}}$, $\vec{B}^{\pm}_{\mbox{\scriptsize{p}}}$ are the contributions of magnetic fields from the
spectator constituents ($\mbox{N}_{\mbox{\scriptsize{s}}}$) and participating constituents ($\mbox{N}_{\mbox{\scriptsize{p}}}$),
moving in the positive (negative) direction of the $Z$ axis. The contributions from spectators and participants can be estimated with help of the following equations \cite{NPA-803-227-2008}
\begin{subequations}
\begin{equation}
e\vec{B}^{\pm}_{\mbox{\scriptsize{s}}}(\tau,\tilde{y},\vec{x}_{\perp})
= \pm \mbox{Z} \alpha_{\mbox{\scriptsize{EM}}} \sinh(y_{0} \mp
\tilde{y})
\int\,d^{2}\vec{x}^{\,'}_{\perp}\rho_{\pm}(\vec{x}^{\,'}_{\perp})\bigl[1-\theta_{\mp}(\vec{x}^{\,'}_{\perp})\bigr]
\vec{\zeta}^{\pm}(\tau,\tilde{y},\vec{x}_{\perp},\vec{x}^{\,'}_{\perp},y_{s}),
\label{eq:2.2a}
\end{equation}
\begin{equation}
e\vec{B}^{\pm}_{\mbox{\scriptsize{p}}}(\tau,\tilde{y},\vec{x}_{\perp}) = \pm \mbox{Z} \alpha_{\mbox{\scriptsize{EM}}} \int\,d^{2}\vec{x}^{\,'}_{\perp}
\rho_{\pm}(\vec{x}^{\,'}_{\perp})\theta_{\mp}(\vec{x}^{\,'}_{\perp})\int_{-y_{0}}^{y_{0}}dy_{\mbox{\scriptsize {p}}}
f(y_{\mbox{\scriptsize{p}}})\sinh(y_{\mbox{\scriptsize{p}}} \mp \tilde{y})\vec{\zeta}^{\pm}(\tau,\tilde{y},\vec{x}_{\perp},\vec{x}^{\,'}_{\perp}, y_{\mbox{\scriptsize {p}}}), \label{eq:2.2b}
\end{equation}
\begin{equation}
\vec{\zeta}^{\pm}(\tau,\tilde{y},\vec{x}_{\perp},\vec{x}^{\,'}_{\perp},\kappa) =
\frac{(\vec{x}^{\,'}_{\perp}-\vec{x}_{\perp}) \times \vec{e}_{\scriptsize{Z}}}{\bigl[(\vec{x}^{\,'}_{\perp}-\vec{x}_{\perp})^{2}+\tau^{2}
\sinh(\kappa \mp \tilde{y})^{2}\bigr]^{3/2}}. \nonumber
\end{equation}\label{eq:2.2}
\end{subequations}
Here $y_{\mbox{\scriptsize {s/p}}}$ is the rapidity of spectator /
participants $\mbox{N}_{\mbox{\scriptsize{s/p}}}$ in the
laboratory reference frame, which is coincide with the
center-of-mass system for collider beams, $\tau =
\sqrt{t^{2}-z^{2}}$ and $\tilde{y} = 0.5\ln[(t+z)/(t-z)]$ is
proper time and space-time rapidity,
$\rho_{\pm}(\vec{x}^{\,'}_{\perp})$ is the constituent density. It
is considered that spectators $\mbox{N}_{\mbox{\scriptsize{s}}}$
do not (re)scatter and continue to move along the $Z$ axis with
$y_{0}$ after the interaction, i.e. $y_{\mbox{\scriptsize
{s}}}=y_{0}$. Within the hypothesis about negligible contribution
of the newly produced particles to the $\vec{B}$
\cite{NPA-803-227-2008} the function $f(y{\mbox{\scriptsize {p}}})
= [a\exp(ay{\mbox{\scriptsize {p}}})]/[2\sinh(ay_{0})]$,
$y{\mbox{\scriptsize {p}}} \in [-y_{0}, y_{0}]$ is entered to
account for contributions only from $\mbox{N}_{\mbox {\scriptsize
{p}}}$ presented in the initial state, where $a \simeq 0.5$ based
on available experimental data \cite{PLB-378-238-1996}.

\begin{figure*}
\includegraphics[width=16.0cm,height=16.0cm]{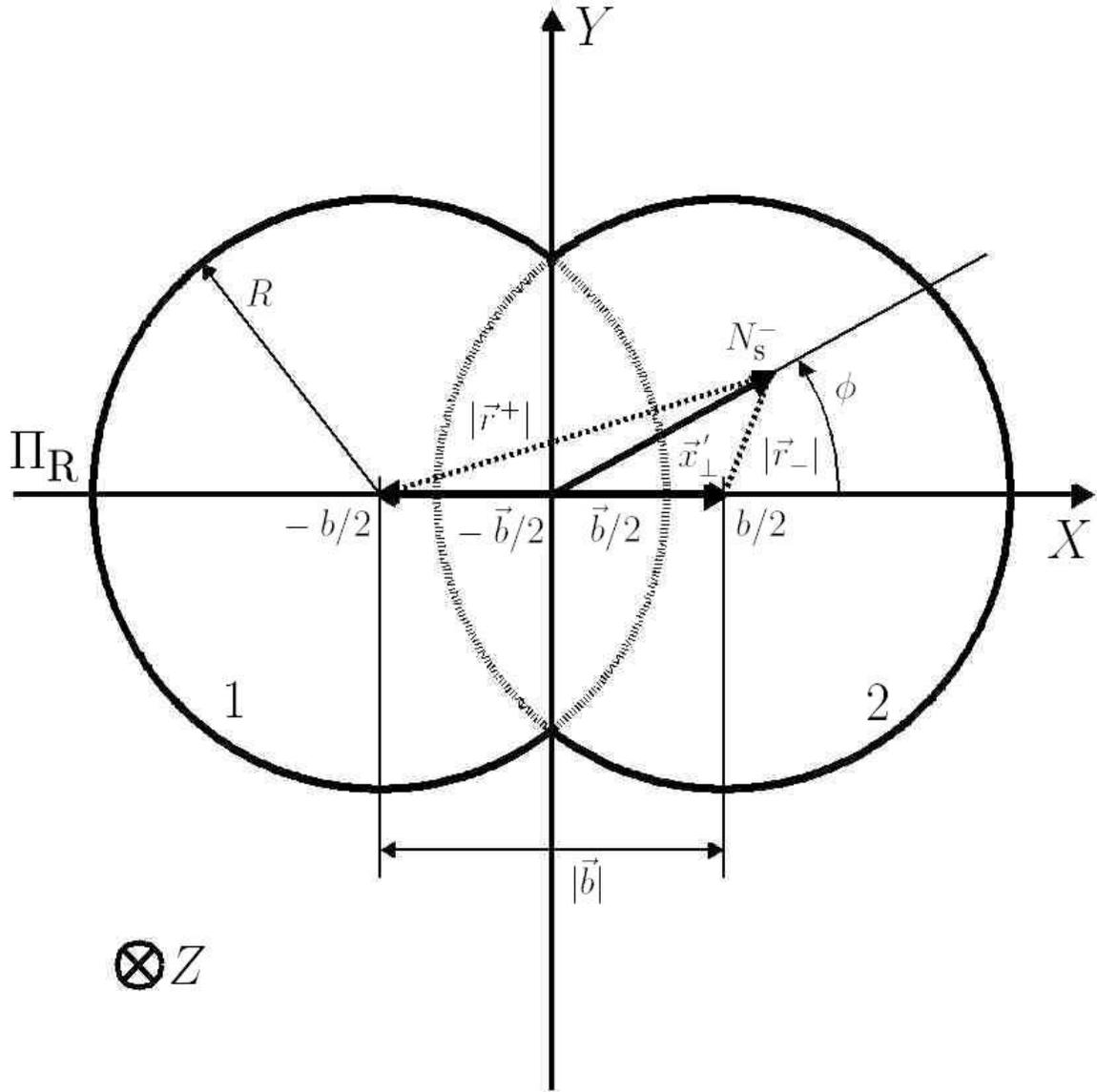}
\vspace*{8pt} \caption{
Detailed picture for collision geometry of two equal objects with finite size (proton / nucleus) in the plane
transverse with respect to the beam axis ($Z$) within hard sphere model. The objects with radii $R$,
moving in opposite directions (the object 1 -- in the
positive direction of $Z$, the object 2 -- in negative), collide
with the impact parameter $|\vec{b}|$. In this case, without lost of
generality, the coordinate system is chosen in such a way that
$XZ$ plane coincides with the reaction plane $\Pi_{\mbox{\scriptsize{R}}}$ and, consequently, the angle $\phi$ is the
angle relative to $\Pi_{\mbox{\scriptsize {R}}}$. Region
the overlap of the two objects, shown by the dotted curve, contains
participating constituents, the spectator constituents are outside the specified area.
In the $XY$ plane, the position of the spectator $N^{-}_{\mbox{\footnotesize{s}}}$ is characterized
by the vector $\vec{x}^{\,'}_{\footnotesize {\perp}}$ relative to
the origin of the coordinate system and by the vectors $\vec{r}^{\,+}_{-}=\vec{x}^{\,'}_{\perp} \pm \vec{b}/2$
shown as the dashed lines with respect to the centers of the objects 1 and 2 respectively.} \label{fig:1}
\end{figure*}

In relativistic energy domain the Lorentz factor
$\gamma_{\footnotesize{N}} \simeq
\sqrt{\smash[b]{s_{\footnotesize{NN}}}}\,/\,2m_{\footnotesize{N}} \gg 1$ and
colliding objects (proton / nucleus) are strongly contracted in
the longitudinal ($Z$) direction of their original size, where $m_{\footnotesize{N}}$ is the nucleon mass \cite{PDG-PRD-98-030001-2018}. Therefore
in the simple approximation "hard sphere'' the constituent density
is defined as $\rho_{\pm}(\vec{x}^{\,'}_{\perp}) = 3\bigl[R^{2}-
(\vec{x}^{\,'}_{\perp} \pm
\vec{b}/2)^{2}\bigr]^{1/2}\,\theta_{\pm}(\vec{x}^{\,'}_{\perp})\,/\,(2\pi
R^{3})$ for the charged object moving in a positive (negative)
direction along the $Z$ axis, where normalization is the following
$\displaystyle \int
d^{2}\,\vec{x}^{\,'}_{\perp})\rho_{\pm}(\vec{x}^{\,'}_{\perp})=1$
and $\theta_{\pm}(\vec{x}^{\,'}_{\perp}) = \theta
\bigl[R^{2}-(\vec{x}^{\,'}_{\perp}\pm\vec{b}/2)^{2}\bigr]$ are the
projections of the colliding objects on the transverse plane with
respect to the beam axis, $\theta(x)$ is the step function used
for splitting $\mbox{N}_{\mbox{\scriptsize{s}}}$ and
$\mbox{N}_{\mbox{\scriptsize{p}}}$ in the approach considered, $R$
is the radius of the beam object (proton / nucleus). Fig.
\ref{fig:1} shows in detailed the collision geometry and
parameters used for calculation of $\vec{B}$ with help of
(\ref{eq:2.1}).

Within the hard sphere approach and in the center of the secondary particle source, i.e. in the center of the overlap region ($|\vec{x}_{\perp}|=0$ and $\tilde{y}=0$), the magnetic field points along the $Y$ axis: $\vec{B} = B\vec{e}_{\scriptsize{Y}}$ \cite{NPA-803-227-2008,Okorokov-arXiv-0908.2522-2009}. This statement agrees well with the averaged results of event-by-event numerical calculations for various nucleus-nucleus collisions. The improvement of estimation of the $\vec{B}_{\mbox{\scriptsize{p}}}^{\pm}$ provides the following analytic approximation \cite{Okorokov-YaFE-4-805-2013}:
\begin{equation}
\displaystyle
eB \simeq 8 \mbox{Z}\alpha_{\mbox{\scriptsize{EM}}}\frac{b}{\tau^{3/2}}\exp(-y_{0}/2)
\biggl[\frac{\tilde{c}f(x)}{xR^{3/2}}+\frac{\exp(-3y_{0}/2)}{\tau^{3/2}}\biggr], \label{eq:2.3}
\end{equation}
which is valid for proper time range $\tau \in [\tau_{1}, \tau_{2}]$ with $\tau_{1} = R\,/\sinh(y_{0})$ and $\tau_{2}=R$. In (\ref{eq:2.3})
the first term corresponds to the contribution of $\mbox{N}_{\mbox{\scriptsize{p}}}$ and the second one -- to the contribution of spectators
$\mbox{N}_{\mbox{\scriptsize{s}}}$, $\tilde{c} \simeq 0.075$,
$x \equiv b/R$, the function $f(x) = \sum_{\pm}\mp \sqrt{R} \displaystyle \int d^{\,2}\vec{x}^{\,'}_{\perp}
\rho_{\pm}(\vec{x}^{\,'}_{\perp})\theta_{\mp}(\vec{x}^{\,'}_{\perp})x^{\,'}|\vec{x}^{\,'}_{\perp}|^{-3/2}$ is calculated numerically \cite{NPA-803-227-2008}. At high energies $y_{0} \gg 1$ and it can be derived the following analytic expressions for limit values of the magnetic filed within the approximation (\ref{eq:2.3}):
\begin{eqnarray}
\displaystyle
(eB)|_{\tau_{2}} &\approx& 8\mbox{Z}\alpha_{\mbox{\scriptsize{EM}}}\frac{\tilde{c}f(x)}{R^{2}}\exp(-y_{0}/2), \nonumber\\
(eB)|_{\tau_{1}} &\approx& (eB)|_{\tau_{2}}\times\sinh^{3/2}y_{0}\biggl[1+\frac{0.5x}{\tilde{c}f(x)}\biggr]. \label{eq:2.4}
\end{eqnarray}

It should be noted that the analytic equations (\ref{eq:2.0}) and
(\ref{eq:2.3}) deduced within point charge and hard sphere
approaches do not take into account any possible modifications due
to matter produced in the final state, i.e. the models used here
corresponds to the $\vec{B}$ in vacuum.

\subsection{\label{subsec:2.2}Beam characteristics}
In the present paper the energies are considered for the following
international projects: the novel research infrastructure based on
the Large Hadron Collider (LHC), which extends the current energy
frontier by almost a factor 2 is called the high-energy Large Hadron Collider (HE--LHC) project
\cite{FCC-CDR-V4-2018} and the integrated accelerator facility in
a global context is called the Future Circular Collider (FCC) project which contains the work mode (FCC--hh) with proton and
nuclei beams \cite{FCC-CDR-V3-2018}. Both projects are essential
part for the next update of the European strategy for particle
physics. The nominal energy for proton--proton collision within
the HE--LHC project is
$\sqrt{\smash[b]{s_{\footnotesize{pp}}}}=27$ TeV
\cite{FCC-CDR-V4-2018} and
$\sqrt{\smash[b]{s_{\footnotesize{pp}}}}=100$ TeV for the FCC
project \cite{FCC-CDR-V3-2018}. Within colliding nuclei
$(\mbox{A}_{1}, \mbox{Z}_{1}) + (\mbox{A}_{2}, \mbox{Z}_{2})$ with
nucleon numbers $\mbox{A}_{1}$, $\mbox{A}_{2}$ and charges
$\mbox{Z}_{1}e$, $\mbox{Z}_{2}e$ in rings with magnetic field set
for protons of momentum $p_{\,0,p}$ and mass $m_{p}$ \cite{PDG-PRD-98-030001-2018}, the colliding nucleon pairs
will have an average beam energy and centre-of-mass energy
\cite{arXiv-1812.06772}
\begin{equation}
\forall\,i=1,2: \left.E_{0,i}\right|_{m_{N} \approx m_{p},\,
m_{p} \ll p_{\,0,p}} \approx
\mbox{Z}_{i}E_{0,p}/\mbox{A}_{i},~~~
\left.\sqrt{\smash[b]{s_{\footnotesize{NN}}}}\right|_{m_{N} \approx m_{p},\,
m_{p} \ll p_{\,0,p}} \approx
\sqrt{\smash[b]{s_{\footnotesize{pp}} \times
(\mbox{Z}_{1}\mbox{Z}_{2}/\mbox{A}_{1}\mbox{A}_{2})}}.
\label{eq:2.5}
\end{equation}

This work devotes the study of symmetric collisions. Some nuclei, from light to heavy, are considered as beam particles for high-luminosity LHC \cite{arXiv-1812.06772}. It seems reasonable for complete information to consider the same nuclei as incoming particles at energies of the HE--LHC and FCC projects. Table \ref{tab:1} shows the essential kinematic parameters for various nuclei, where the first line corresponds to the $\sqrt{\smash[b]{s_{\footnotesize{pp}}}}=27$ TeV and second line -- to the $\sqrt{\smash[b]{s_{\footnotesize{pp}}}}=100$ TeV for each parameter considered.

\begin{table*}
\caption{\label{tab:1}Kinematic parameters for various beams.}
\begin{center}
\begin{tabular}{cccccccc}
\hline \multicolumn{1}{c}{Parameter} &
\multicolumn{7}{c}{Incoming particle} \rule{0pt}{10pt}\\
\cline{2-8}
 & ${}^{1}p^{1+}$ & ${}^{16}\mbox{O}^{8+}$ & ${}^{40}\mbox{Ar}^{18+}$ & ${}^{40}\mbox{Ca}^{20+}$ & ${}^{78}\mbox{Kr}^{36+}$ & ${}^{129}\mbox{Xe}^{54+}$ & ${}^{208}\mbox{Pb}^{82+}$ \rule{0pt}{10pt}\\
\hline
$E_{0}$, TeV & 13.50 & 6.750 & 6.075 & 6.750 & 6.231 & 5.651 & 5.322 \\
             & 50.00 & 25.00 & 22.50 & 25.00 & 23.08 & 20.93 & 19.71 \\
$y_{0}$      & 10.31 & 9.576 & 9.472 & 9.576 & 9.492 & 9.396 & 9.341 \\
             & 11.51 & 10.82 & 10.77 & 10.82 & 10.58 & 10.73 & 10.70 \\
\hline
\end{tabular}
\end{center}
\end{table*}

The radius for beam particle is estimated as the radius of
spherically-symmetric object $\forall\,A: R = r_{0}A^{1/3}$ with
$r_{0}=(1.25 \pm 0.05)$ fm \cite{book-1,book-2}. For $p+p$
interactions the quark-quark collisions ($q+q$) can be also
considered with the following estimations
$\sqrt{\smash[b]{s_{\footnotesize{qq}}}} \sim
\sqrt{\smash[b]{s_{\footnotesize{pp}}}}/3$
\cite{arXiv-hep-ph-0410324-2004} and $r_{q} \sim
R_{p}/3$ for constituent quarks.

The approach of the point-like charge can be applied for the finite-size object with characteristic linear scale (radius) $R$ at sufficiently large impact parameters $b \gg R$. The amplitude value of magnetic field from (\ref{eq:2.0}) can be re-written $eB_{0} \propto Z/x^{2}R^{2}$. Thus, one can see  from (\ref{eq:2.0}) and (\ref{eq:2.4}) the amplitude $eB_{0}$ and extremes $(eB)|_{\tau_{1}}$, $(eB)|_{\tau_{2}}$ of the hard-sphere model show the similar behavior $X \propto Z/R^{2}$, $X \equiv eB_{0}$, $(eB)|_{\tau_{1}}$ or $(eB)|_{\tau_{2}}$ with changes of the charge number and radius of beam particle and at fixed $x$. The following approximate empirical relation is valid for an isobar of a selected nucleus that is stable relative to $\beta$-decay \cite{book-2}: $\displaystyle Z \approx A/(1.980+0.015A^{2/3})$. Then, one can derive the $Z$- and $A$-dependence of the quantity $X$ within point-like particle and hard-sphere approaches for magnetic field
\begin{eqnarray}
\displaystyle
X(Z) &\propto& Z, \nonumber\\
X(A) &\propto& A^{1/3}/(1.980+0.015A^{2/3}). \label{eq:2.6}
\end{eqnarray}

\section{Results}\label{sect:3}

Numerous phenomenological studies are devoted to the
(electro)magnetic fields arising from the nucleus-nucleus collisions\footnote{See, for instance, some review papers
\cite{AHEP-2013-490495-2013,RPP-79-076302-2016} and references therein. The overview of electromagnetic probe production in heavy ion collisions at relativistic energies can be found elsewhere \cite{JPCF-832-012035-2017}.}. However, up to now the
papers are for the energies
$\sqrt{\smash[b]{s_{\footnotesize{NN}}}} < 7$ TeV \cite{AHEP-2014-193039-2014} and they are
usually focused on the heavy ion collisions. In this work, the
strength of external magnetic field is estimated within approaches
of point-like charges (\ref{eq:2.0}) and hard sphere for
collisions of the particles from Table \ref{tab:1} with
$\sqrt{\smash[b]{s_{\footnotesize{NN}}}}$ corresponding to the
nominal proton-proton collision energies within HE--LHC and
FCC--hh projects for the first time.

The $B(t)$ dependence at the center of collision obtained within
the approach of point-like charges shows rapid decrease with $t$
for any beam types from Table \ref{tab:1}, especially for $p+p$.
The (\ref{eq:2.0}) allows the estimations for amplitude of the
magnetic field and characteristic time depends on the beam type
and $\sqrt{\smash[b]{s_{\footnotesize{NN}}}}$. The results for
amplitude are following $eB_{0} \sim 20$ GeV$^{2}$ for $p+p$ and
this parameter is in the range $\sim (13-19)$ GeV$^{2}$ for rest
nuclei; $t_{0} \sim 0.4 \times 10^{-4}$ fm/$c$ for $p+p$ and
$(2-7) \times 10^{-4}$ fm/$c$ for other nuclei at
$\sqrt{\smash[b]{s_{\footnotesize{pp}}}}=27$ TeV. The
corresponding estimations are $eB_{0} \sim 78$ GeV$^{2}$ for $p+p$
and $\sim (49-71)$ GeV$^{2}$ for rest of Table \ref{tab:1}; $t_{0}
\sim 0.1 \times 10^{-4}$ fm/$c$ for $p+p$ and $(0.6-1.8) \times
10^{-4}$ fm/$c$ for other nuclei at
$\sqrt{\smash[b]{s_{\footnotesize{pp}}}}=100$ TeV. Values of the
parameters $eB_{0}$ and $t_{0}$ are mostly growth for transition
from $\mbox{O}+\mbox{O}$ to $\mbox{Pb}+\mbox{Pb}$ collisions. The
quantitative results above correspond to the semi-central
collisions ($x=1$).

Figs. \ref{fig:2}, \ref{fig:3} show the dependence $eB(\tau)$ for
$p+p$ (a), $\mbox{O}+\mbox{O}$ (b), $\mbox{Xe}+\mbox{Xe}$ (c) and
$\mbox{Pb}+\mbox{Pb}$ (d) collisions in central (dashed lines),
semi-central (solid curves) and peripheral (dotted lines) events
at at $\sqrt{\smash[b]{s_{\footnotesize{pp}}}}=27$ and 100 TeV
respectively. These smooth curves are obtained within hard sphere
model for the range $R\,/\sinh(y_{0}) \lesssim \tau \lesssim R$
with help of the analytic equation (\ref{eq:2.3}). The magnetic
field strength for particle species from Table \ref{tab:1} at
HE--LHC (Fig. \ref{fig:2}) and FCC--hh (Fig. \ref{fig:3}) energies
decreases fast with $\tau$ increase in a vacuum, especially for
peripheral collisions. Such behavior of the $eB(\tau)$ dependence
is in agreement with previous results at lower energies
\cite{Okorokov-arXiv-0908.2522-2009,Okorokov-YaFE-4-805-2013}. On
the left boundaries of the temporary ranges studies the absolute
value of magnetic filed reaches the extremely large values which
are in order of magnitude $eB \sim 10$ (30) GeV$^{2}$ at
$\sqrt{\smash[b]{s_{\footnotesize{pp}}}}=27$ (100) TeV depending
on the type of beam and centrality. The estimations for $eB$
derived within the hard sphere approach for $\mbox{Au}+\mbox{Au}$
collisions at $\sqrt{\smash[b]{s_{\footnotesize{NN}}}}=0.2$ TeV
\cite{Okorokov-YaFE-4-805-2013} reasonably agree with the results
from the UrQMD \cite{IJMPA-24-5925-2009} and HIJING
\cite{PRD-85-044907-2012} models, whereas the amplitude value of $B$
depends more weakly on centrality, and $eB(\tau)$ dependence
decreases faster with $\tau$ in last cases than that for hard sphere
approach. Furthermore, there is a similar situation between the
hard-sphere results and calculations from HSD model
\cite{PRC-83-054911-2011} in which some smaller value of amplitude of
$B$ and faster decrease of magnetic field strength with increase
of $\tau$ is predicted with respect to the corresponding results
obtained with help of (\ref{eq:2.3}). The reasonable agreement
between various approaches proves that the hard-sphere model can
be considered as appropriate approach for estimation of the time
evolution of (electro)magnetic field strength in particle
collisions. Therefore with taking into account much higher
energies studied here it allows the qualitative expectation that
the equation (\ref{eq:2.3}) provide the reasonable estimations for
$eB(\tau)$ in HE-LHC and FCC-hh energy domains. The previous
analysis \cite{Okorokov-YaFE-4-805-2013} shown also that
$(eB)|_{\tau_{1}} \simeq (eB)_{\mbox{\scriptsize{max}}}$ at least
for $\sqrt{\smash[b]{s_{\footnotesize{NN}}}}=0.1$ TeV. Moreover
numerical calculations within various approaches predicted the
peak in the dependence $eB(\tau)$ shown that the width of the peak
decreases with growth $\sqrt{\smash[b]{s_{\footnotesize{NN}}}}$.
Those one can expect that the values $(eB)|_{\tau_{1}}$ obtained
here for ultra high energies (Figs. \ref{fig:2}, \ref{fig:3}) are
reasonable estimations for amplitude values of the strength of
magnetic field in nuclear collisions. The weakest dependence both
on the beam type and on $\sqrt{\smash[b]{s_{\footnotesize{NN}}}}$
is observed for peripheral collisions, in which a dominant
contribution to $\vec{B}$ comes from spectator nucleons. At fixed
$\tau$ (i) the magnetic-field strength is larger for collisions of
heavier nuclei; (ii) the magnetic field becomes weaker as
$\sqrt{\smash[b]{s_{\footnotesize{NN}}}}$ grows within range of
time $R\,/\sinh(y_{0}) \lesssim \tau \lesssim R$ accessible for
the analytic equation (\ref{eq:2.3}). The last effect is due to a
faster divergence of spectator nucleons at the increase of
$\sqrt{\smash[b]{s_{\footnotesize{NN}}}}$, whose contribution
depends strongly on the rapidity of beam particles. These
relations between curves shown in Figs. \ref{fig:2}, \ref{fig:3}
for various beam types and
$\sqrt{\smash[b]{s_{\footnotesize{NN}}}}$ coincide with results of
the previous works
\cite{Okorokov-arXiv-0908.2522-2009,Okorokov-YaFE-4-805-2013}.

Here the proton radius is estimated in accordance with the general
approach $R \propto A^{1/3}$ for all elements from the periodic
table. On the other hand such method provides significant
overestimation of the $R_{p}$ with respect to the "preferable"
CODATA value $R_{ch,p}^{\,ep~\scriptsize{\mbox{CODATA}}}= 0.875 \pm
0.006$ fm for charge radius of the proton
\cite{Okorokov-IJMPA-33-1850077-2018}. As consequence the
overestimation leads to the decrease of the first term in
(\ref{eq:2.3}) and the increase of $\tau_{1}$. This additional
uncertainty is understandable for proton and corresponding study
is in the progress. As observed at smaller collision energies a
consistent transition from the simplest approximation of
point-like sources to the hard sphere model and the two-component
Fermi model
\cite{Okorokov-JPConfSer-668-012129-2016,Okorokov-JPConfSer-675-022021-2016}
for the nucleon-density distribution in a nucleus leads to a
decrease in $B_{\scriptsize{\mbox{max}}}$, especially for the
first two models \cite{Okorokov-PAN-80-1133-2017}. It seems the
agreement between model with point-like charges and hard sphere
model is some better at ultra high energies of the HE--LHC and
FCC--hh projects. Those one can expect some decrease the
$eB(\tau)$ values for more accurate Fermi model with respect to
those presented here but possibly this changing will not be
dramatic. The calculations are in the progress for the magnetic
field in ultra-high energy particle collisions with Fermi model
for the nucleon distributions nucleus.

Also the time dependence of $B$ is studied for $q+q$ collisions within the approach of point-like charges. Based on the (\ref{eq:2.0}) the following estimations are obtained for values of the magnetic-field amplitude and characteristic time: $eB_{0} \sim 0.6\,(2.3) \times 10^{2}$ GeV$^{2}$ and $t_{0} \sim 1.4\,(0.4) \times 10^{-5}$ fm/$c$ for the $\mbox{Z}_{q}=1/3$ corresponds to the $d$-quark and $\sqrt{\smash[b]{s_{\footnotesize{pp}}}}=27$ (100) TeV. These estimations correspond to the relative impact parameter $x=1$.

\section{Discussion}\label{sect:4}

As discussed in \cite{Okorokov-PAN-80-1133-2017} collisions of relativistic nuclei also generate very
strong $\vec{E}$. These electromagnetic fields may have a substantial effect
on multiparticle-production processes in quantum chromodynamics (QCD). The hydrodynamic
properties of strongly coupled quark-gluon plasma (sQGP), together with the chiral QCD anomaly and an
extremely strong external magnetic field, lead to the emergence of
anomalous hydrodynamic phenomena, which are manifestations of the
non-Abelian quantum nature of QCD \cite{PPNP-88-1-2016}.
Allied phenomena include currents flowing along the direction of
the magnetic field or inner vorticity. Experimental signatures of
such macroscopic manifestations of the chiral QCD anomaly are
observed in nucleus--nucleus collisions as the separation of electric
charges etc.

\begin{figure*}
\includegraphics[width=16.0cm,height=16.0cm]{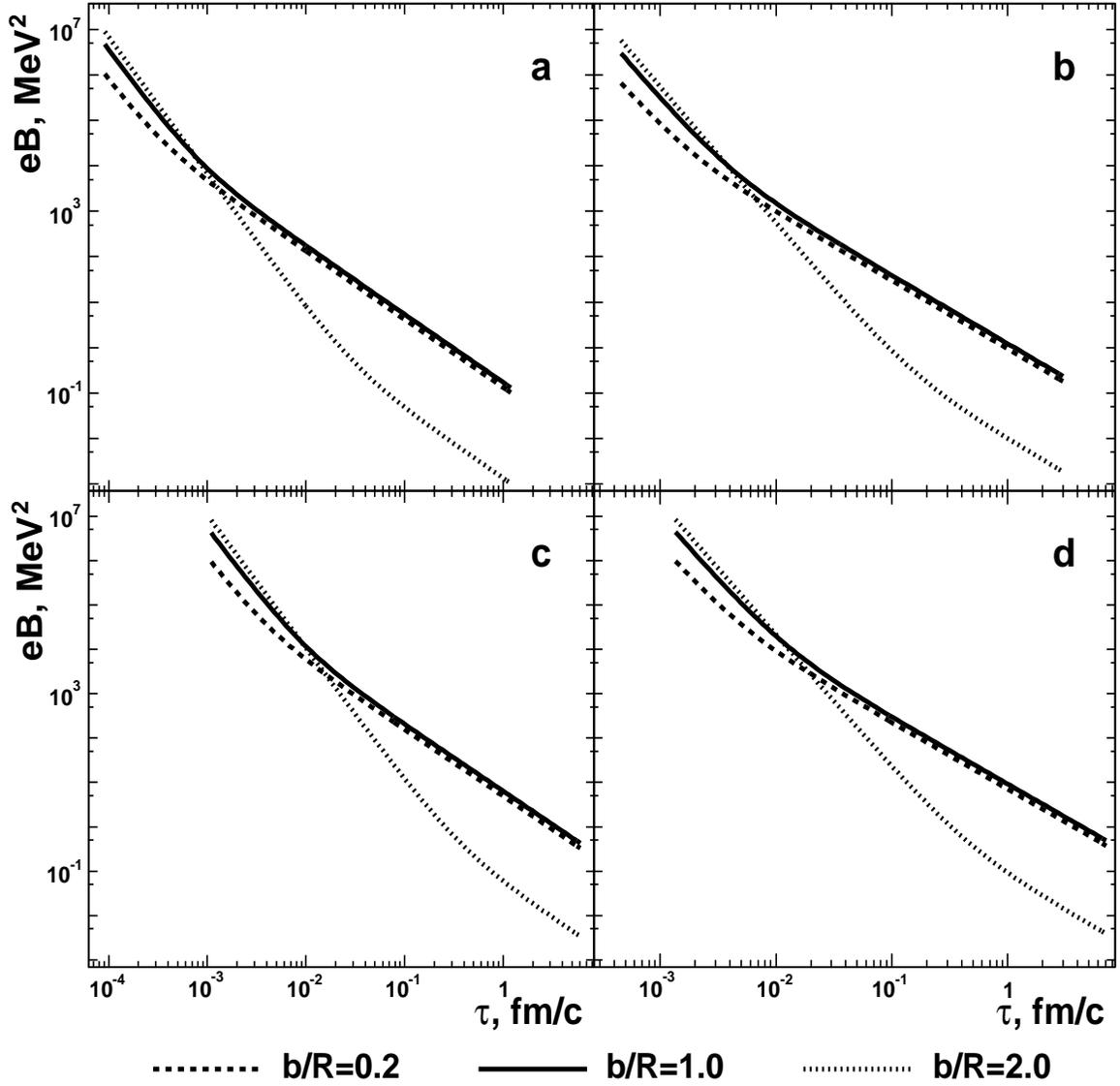}
\vspace*{8pt} \caption{Dependence $eB(\tau)$ calculated with help
(\ref{eq:2.3}) in the range $R\,/\sinh(y_{0}) \lesssim \tau \lesssim R$ for
$p+p$ (a), $\mbox{O}+\mbox{O}$ (b), $\mbox{Xe}+\mbox{Xe}$ (c) and $\mbox{Pb}+\mbox{Pb}$ (d) collisions
at nominal value $\sqrt{\smash[b]{s_{\footnotesize{pp}}}}=27$ TeV for the HE--LHC project. Dashed curves correspond to the central
collisions, solid ones -- semi-central, and dotted curves are for peripheral interactions.} \label{fig:2}
\end{figure*}

\begin{figure*}
\includegraphics[width=16.0cm,height=16.0cm]{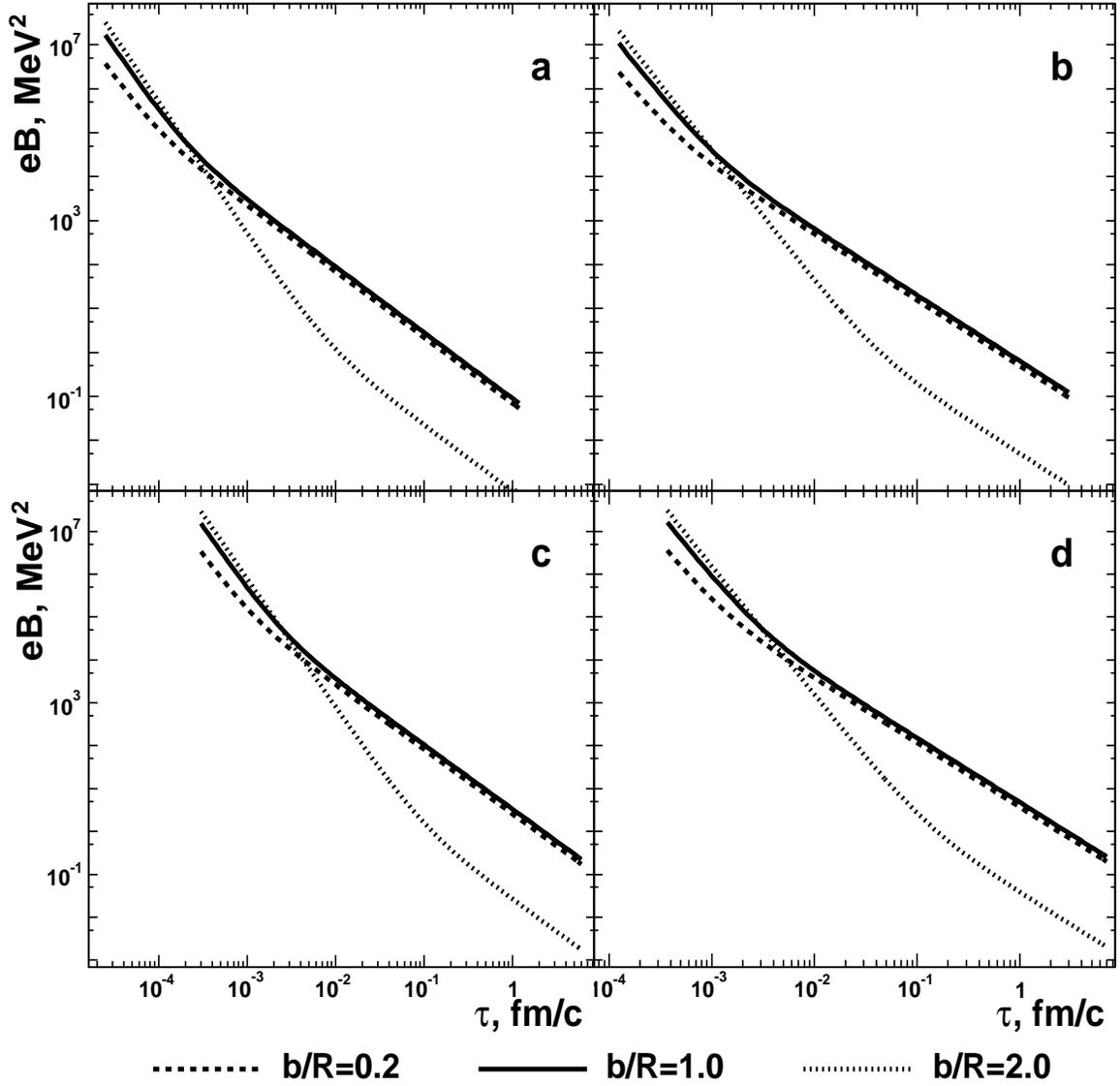}
\vspace*{8pt} \caption{Dependence $eB(\tau)$ calculated with help
(\ref{eq:2.3}) in the range $R\,/\sinh(y_{0}) \lesssim \tau \lesssim R$ for
$p+p$ (a), $\mbox{O}+\mbox{O}$ (b), $\mbox{Xe}+\mbox{Xe}$ (c) and $\mbox{Pb}+\mbox{Pb}$ (d) collisions
at nominal value $\sqrt{\smash[b]{s_{\footnotesize{pp}}}}=100$ TeV for the FCC--hh project. Notations used for the smooth curves are the same as in Fig. \ref{fig:2}.} \label{fig:3}
\end{figure*}

One can note that $eB$ reaches the value on about $10^{2}$ MeV in
central and semi-central $p+p$ collisions for $\tau \sim 0.1$
fm/$c$ at energy of HE--LHC (Fig. \ref{fig:2}a) and FCC-hh (Fig.
\ref{fig:3}a) project. This value corresponds in order of
magnitude to the range of low boundary for the strength of
magnetic field at which experimental manifestation of chiral
magnetic effect (CME) appears
$(eB)_{\mbox{\scriptsize{min}}}^{\mbox{\scriptsize{CME}}} \sim
(\alpha_{S}T)^{2} \sim 10^{2}-10^{3}$ MeV$^{2}$
\cite{NPA-803-227-2008}. One other hand the $\tau \sim 0.1$ fm/$c$
agrees reasonably with the estimations for the onset the
thermalization of the glasma into a sQGP. Moreover the
magnetic-field lifetime increases dramatically upon taking into
account the conductivity of matter and its expansion
\cite{AHEP-2013-490495-2013}. Therefore the present investigation
of the magnetic field within hard sphere model indicates that the
HE--LHC and FCC--hh projects can provide the novel possibility for
study the chiral effects, for instance, CME in $p+p$ collisions.
One can expects the background effects in $p+p$ events will be
significantly weaker than that in nucleus-nucleus collisions at
the same energy. Also extremely large values of $B$ at ultra high
energies and high luminosities of HE--LHC and FCC--hh can provide
the opportunity for study of flavor dependence of the $\mathcal{P
/ CP}$ violation with help the azimuthal correlations of various
particle species. Thus experimental study of topology of QCD
vacuum can be one of the focuses for studies of bulk properties at
the HE--LHC, FCC. Furthermore, the $eB \gtrsim 10^{-2}$ GeV$^{2}$
would also have a profound effect on the breaking of the $SU(2)
\times SU(2)$ chiral symmetry of the strong interactions
\cite{PR-348-163-2001}. As seen in Figs. \ref{fig:2}, \ref{fig:3},
the strength of the magnetic field achieves such values within the
hard sphere model at $\tau \lesssim 10^{-3}$ fm/$c$ in $p+p$
(Figs. \ref{fig:2}a, \ref{fig:3}a) and $\mbox{O}+\mbox{O}$ (Figs.
\ref{fig:2}b, \ref{fig:3}b) interactions and at significantly larger times
$\tau \lesssim 10^{-2}$ in collisions of heavier ions
$\mbox{Xe}+\mbox{Xe}$ (Figs. \ref{fig:2}c, \ref{fig:3}c),
$\mbox{Pb}+\mbox{Pb}$ (Figs. \ref{fig:2}d, \ref{fig:3}d). Thus the
extremely strong magnetic field can affect chirality during the
non-equilibrium very early stages of space-time evolution of the
final-state strongly interacting matter.

The HE--LHC, FCC--hh facilities open the novel opportunity for
study of polarization phenomena in hot environment in particular
the precise measurements of the difference in polarization of
primary $\Lambda$ and $\bar{\Lambda}$ and polarization of heavier
hyperons (for instance, $\Sigma$). Considered extremely strong $eB
\sim 10$ GeV$^{2}$ will provides important changes for behavior of
quarkonium and, perhaps, more heavier particles and states in
particular $\bar{t}t$. Also electromagnetic fields created in
particle collisions at ultra relativistic energies supposed within
HE--LHC, FCC--hh projects can be useful for the study of
fundamental properties of theory namely non-linear or
non-commutative features of quantum electrodynamics (QED), e.g., with help of the light-by-light scattering, production
the magnetic monopoles by the electromagnetic dual of Schwinger
pair creation, etc. \cite{Okorokov-arXiv-1812.07688-2018}. However,
these semi-qualitative suggestions should be verified by
additional quantitative and detailed analysis.

In \cite{Okorokov-AHEP-2016-5972709-2016} the possible effect of
Bose--Eistein condesation in $p+p$ and nucleus--nucleus collisions
at FCC--hh energies was considered in detailed. The closely
related topic is the study of influence of the very short pulse of
the extremely strong Abelian (electro)magnetic field on the
particle production, in particular, pion condensation in external
field \cite{PRL-120-032001-2018}. Furthermore, as shown above the
amplitude value of the magnetic field expected for $q+q$
collisions is $eB_{0} \sim 1.0~(4.0) \times 10^{22}$ G for the
$\mbox{Z}_{q}=1/3$ corresponds to the $d$-quark and
$\sqrt{\smash[b]{s_{\footnotesize{pp}}}}=27$ (100) TeV. These
values are close in order to magnitude to the estimation for
strength of $\vec{B}$ at which $W$ boson condensation occurs
\cite{PLB-257-201-1991}. The amplitude values of $eB$ for any
considered interactions ($A+A$ from Table \ref{tab:1} and $q+q$
collisions) are in the range of so called very intense magnetic
fields ($10^{18}$ -- $10^{24}$ G) which can be generated in the
early Universe \cite{PR-348-163-2001}. Such $B$ can influence the structure of the electroweak vacuum and on the properties of
corresponding phase transition. Therefore, the extremely strong $B$
at HE--LHC and FCC--hh energies can influence on the electroweak
processes.

\section{Conclusions}\label{sect:5}
Summarizing the foregoing, one can draw the following conclusions.

Collisions of relativistic particles are a source of the strongest
electromagnetic field known in nature. For the first time the
estimations for absolute value of magnetic field are obtained
within various approaches for proton and nuclear beams at ultra
high energies corresponded to the HE--LHC and FCC--hh projects.
The analytic approaches used for estimation of strength of the
magnetic field do not take into account the possible influence of
the matter created in final state, i.e. the approaches correspond
to the vacuum conditions. The model with point-like charges
predicts the peak value of $eB$ about $(13-20)$ GeV$^{2}$ at
$\sqrt{\smash[b]{s_{\footnotesize{pp}}}}=27$ TeV and $(49-71)$ at
$\sqrt{\smash[b]{s_{\footnotesize{pp}}}}=100$ TeV while the more
realistic hard sphere approach provides $eB \sim 10$ and 30
GeV$^{2}$. The strength of magnetic field rapidly decrease with
time and increases with growth of atomic number. The amplitude for
$eB$ is estimated at level $60$ $(200)$ GeV$^{2}$ in quark-quark
collisions for charges correspond to $d$-quark at nominal
$\sqrt{\smash[b]{s_{\footnotesize{pp}}}}=27$ (100) TeV.

The extremely strong (electro)magnetic field expected at HE--LHC and FCC--hh can influence on strong and electroweak interaction processes. In particular the principle possibilities appear for study of chiral magnetic effect in proton-proton collisions, for $W$ boson condensation and for manifestation of non-commutative features of the quantum electrodynamics. Further development of theoretical and experimental methods are
of crucial importance for drawing more definitive conclusions for these qualitative suggestions.

\section*{Acknowledgments}

This work was supported partly by NRNU MEPhI Academic Excellence
Project (contract No 02.a03.21.0005, 27.08.2013).

\newpage

\newpage

\newpage

\end{document}